\documentclass[10pt,twocolumn]{IEEEtran}
\usepackage[margin=1in]{geometry}

\usepackage{amsmath,amssymb,amsthm,graphicx,cite,hyperref}
\usepackage[caption=false,font=footnotesize]{subfig}
\usepackage{xcolor}
\usepackage{booktabs}

\theoremstyle{plain}

\theoremstyle{definition}

\theoremstyle{remark}

\begin{document}
\title{Neural Network State-Space Estimators}

\author{Minxing~Sun\thanks{Student author: M.~Sun (sunminxing20@gmail.com).},
        Li~Miao,
        Qingyu~Shen,
        Yao~Mao~\IEEEmembership{Senior Member,~IEEE}\thanks{*Corresponding author: Y.~Mao (maoyao@ioe.ac.cn).},
        and~Qiliang~Bao%
\thanks{All authors are with the State Key Laboratory of Optical Field Manipulation Science and Technology, the Institute of Optics and Electronics, Chinese Academy of Sciences, 
Chengdu~610209, China.}%
\thanks{All authors are with the Key Laboratory of Optical Engineering, the Institute of Optics and Electronics, Chinese Academy of Sciences, 
Chengdu~610209, China.}%
\thanks{All authors are also with the School of Electronic, Electrical and Communication Engineering,
University of Chinese Academy of Sciences, Beijing~101408, China.}%
}

\maketitle

\begin{abstract}
Classical state estimation algorithms rely on predefined target's state-space model, which complicates model derivation and limits adaptability when system dynamics change. Neural network based estimators offer a data-driven alternative, but rarely fuse classical estimation theory into their structure and demand large, pre-computed training sets. To overcome these limitations, we propose a unified state-space structure without target's state-space model and treats both the input-layer activations and all network weights as latent states to be estimated online. We instantiate this nonlinear model with three canonical estimators—the extended Kalman estimator, the unscented Kalman estimator, and the particle estimator—to simulate different neural network and demonstrate its generality. We then benchmark our approach against seven leading neural network estimators across three representative scenarios. Results show that our neural network state-space estimators not only retain the robust learning capability, but also match or exceed the accuracy of both classical and pre-trained neural network methods. (Code, data, and more result: \href{https://github.com/ShineMinxing/PaperNNSSE.git}{github.com/ShineMinxing/PaperNNSSE.git})
\end{abstract}


\maketitle

\section{Introduction}\label{sec1}

State estimation algorithms are extensively utilized across various engineering disciplines \cite{TB1990,EX2021,OS1994}. For example, in electro-optical tracking systems \cite{SK2018, AD2023}, they are essential for mitigating optical path disturbances caused by atmospheric turbulence \cite{SC2011}, compensating for delays in hardware and image processing pipelines \cite{RL2022}, and filtering out signal noise generated by sensors and target detection mechanisms \cite{MS2024}. The fundamental principle of these algorithms comprises three stages: predicting the next state using the current estimate and target's state-space model, predicting the corresponding observation via the observation equation, and correcting the state estimate based on the deviations between the actual observation and its prediction.

Estimators can be categorized based on how they establish the relationship between observation deviations and state deviations. One category comprises methods such as the Extended Kalman Estimator (EKE), which computes partial derivatives from the state-space model. The primary limitation of EKE is the complexity involved in deriving analytical solutions for these partial derivatives, particularly in highly nonlinear models \cite{NK1960,MB1997,EB2001,KG2015,VQ2018}. Another category includes the Unscented Kalman Estimator (UKE), which approximates the state distribution by assigning predefined deviations to each estimated state and evaluating the resulting variations in observations. However, UKE faces challenges in accurately capturing the relationships between coupled estimated states and observations \cite{NJ1997,UW2000,UV2000,UJ2004,RZ2019,NX2019,KF2020}. The third category is represented by Particle Estimator (PE), which introduces random perturbations to the estimated states and assesses the resulting observation variations. While PE is better suited for handling nonlinearities, its main drawback is the high computational cost associated with the large number of particles required, which can impede real-time performance \cite{NG1993,PF2001,TA2002,KR2003}.

All three estimation methods rely on predefined state-space models, and even the robust estimators designed to mitigate model inaccuracies can only compensate for a limited range of errors \cite{RG2003, RS2020, RL2022}. In practical applications, defining an accurate state-space model for the target system is often inherently challenging, particularly when tracking non-cooperative targets.

The rapid advancement and widespread adoption of artificial neural network methodologies have profoundly impacted time-domain signal processing, enabling significant breakthroughs in estimation and prediction tasks \cite{ RA2017a, SA2018, AF2021, PK2021, AF2024, PT2024}. Recurrent Neural Network (RNN), introduced by Hopfield \cite{NH1982}, were enhanced by Rumelhart, Hinton, and Williams \cite{LR1986} with the backpropagation algorithm, facilitating more efficient training. Subsequent improvements by Jordan and Elman \cite{SJ1997, FE1990} incorporated feedback loops, enhancing the ability to model temporal dependencies \cite{RM2021, SB2021, RK2021b}. However, RNNs struggled with vanishing and exploding gradients, challenges addressed by Hochreiter \cite{LH1997} with Long Short-Term Memory (LSTM) networks, and further refined by Schmidhuber \cite{LG1999} through the addition of forget gates \cite{SW2020, VI2021}. Gated Recurrent Unit (GRU), proposed by Kyunghyun \cite{LC2014a} and validated by Junyoung \cite{EC2014a}, offered a simplified yet effective alternative to LSTM \cite{MA2020, MG2022, CG2024}. Convolutional Neural Network (CNN), initially developed by LeCun \cite{BL1989} for handwritten digit recognition, have been extended to Recurrent Convolutional Neural Networks \cite{RL2015} to capture spatial and temporal dependencies, significantly improving prediction accuracy in applications such as vehicle and human trajectory forecasting \cite{HM2020, IM2020, NZ2021, AY2024}. Temporal Convolutional Network (TCN), introduced by Colin \cite{TL2017}, utilize pooling and upsampling to capture long-term motion patterns, enhancing capabilities in action segmentation and detection \cite{SC2020, SZ2022a, MW2023, VY2024} (Chen, 2020; Kaiyu, 2022). Neural Ordinary Differential Equation (NeuralODE), proposed by Ricky \cite{NC2018}, represent neural network layers through differential equations, facilitating deeper and more efficient networks for tasks like robotic path planning and travel time estimation \cite{NP2022, TC2023, SK2024}. Transformer architectures, introduced by Ashish \cite{AV2017}, leverage self-attention mechanisms to model long-term dependencies while reducing computational complexity. Their applications span maneuvering target tracking, autonomous vehicle trajectory prediction, and 3D human pose estimation \cite{EQ2022, TZ2023, RS2023, KP2024, TO2024}. 

Traditional estimation algorithms have also been integrated into neural networks to optimize training processes. Early efforts by Singhal, Puskorius, and Pérez-Ortiz \cite{TS1988, NP1994, KP2003, ED2016a} applied EKE to train multilayer perceptrons and RNN, enhancing training efficiency. More recent integrations of UKE and PE have improved the training of feedforward and recurrent networks in applications such as voice classification and evapotranspiration estimation \cite{TD2017, PN2020}. However, in these studies, estimation algorithms primarily treated the backpropagation algorithm as state-space equations, serving an auxiliary role rather than being the primary mechanism for state estimation. Consequently, the effectiveness of estimation algorithms remains limited and warrants further enhancement\cite{IA2023}.

To address these limitations, our proposed algorithm introduces a novel state-space model that treats input layer nodes and all weight parameters as estimated states, effectively simulating a neural network within a state estimation framework. The main contributions of this work are as follows:

\begin{itemize}
  
\item Neural Network State-Space Model (NNSSM): We propose a NNSSM that simulates a neural network, enabling real-time learning and pretraining. The estimation algorithm works wothout target's state-space models and can be tailored by adjusting the input layer length and prediction steps to meet engineering requirements.

\item Neural Network State-Space Estimator (NNSSE): The versatility of the proposed NNSSM is validated through the implementation of three representative estimation algorithms—EKF, UKE, and PF—demonstrating its broad applicability.

\item Flexible Architecture Configuration: The NNSSM can accommodate arbitrary neural network structures—adjusting the number of layers, the number of neurons per layer, connectivity patterns, and activation functions—to meet diverse application requirements.

\item Comparative Performance Evaluation: We compare the estimation performance of our NNSSEs with various neural network architectures, including RNN, LSTM, GRU, TCN, NeuralODE, and Transformer.

\item Comprehensive Testing: Extensive tests were conducted on generated sine trajectories, a laboratory dual-reflection mirror platform, and actual unmanned aerial vehicle trajectory data, demonstrating the robustness and effectiveness of our proposed method.

\end{itemize}

\section{Problem Statement}\label{sec2}
\noindent
In many real-world tracking problems, the target is non-cooperative, so an accurate state-space model cannot be specified a priori.  A common expedient is the uniformly accelerated state-space model in~\eqref{1_4}, where only position can be observed while velocity and acceleration remain hidden:  
\begin{equation}
\label{1_4}
\begin{aligned}
\mathbf x_{i+1} &= 
\begin{bmatrix}
1 & T & \tfrac{T^{2}}{2}\\[2pt]
0 & 1 & T\\
0 & 0 & 1
\end{bmatrix} \mathbf x_i \;+\;
\begin{bmatrix}
1&0&0\\
0&1&0\\
0&0&1
\end{bmatrix} \mathbf u_i,\\[4pt]
\mathbf z_i &= 
\begin{bmatrix}
1&0&0
\end{bmatrix} \mathbf x_i + \mathbf v_i .
\end{aligned}
\end{equation}

\noindent

\medskip
The state vector \(\mathbf x_i=[p_i,v_i,a_i]^{\mathsf T}\) stacks position, velocity and acceleration at instant~\(i\). \(T\) is the time interval between successive estimates. The process noise \(u_i\) absorbs modelling errors and environmental perturbations, and the measurement noise \(v_i\) reflects sensor inaccuracies; both are assumed zero-mean Gaussian and seperately follow the covariance \(\mathbf Q\), \(\mathbf R\).  Their second-order statistics together with the initial error covariance \(\mathbf \Pi_{0|0}\) are summarised in~\eqref{1_2}, in which \(\delta_{ij}\) denotes the Kronecker delta:  

\begin{equation}
\label{1_2}
E\!\left(
\begin{bmatrix}\mathbf x_0\\\mathbf u_i\\\mathbf v_i\end{bmatrix}
\begin{bmatrix}\mathbf x_0\\\mathbf u_j\\\mathbf v_j\end{bmatrix}^{\!\mathsf T}
\right)=
\begin{bmatrix}
\mathbf \Pi_{0|0} & 0 & 0\\
0 & \mathbf Q\,\delta_{ij} & 0\\
0 & 0 & \mathbf R\,\delta_{ij}
\end{bmatrix}.
\end{equation}

\noindent

\medskip
Because the true motion can be highly nonlinear, model~\eqref{1_4} rarely yields the desired predictive accuracy.  Although the Kalman estimator provides on-line estimates of \(v_i\) and \(a_i\), they are ultimately reconstructed from past position measurements.  With scalar gains \(w_{v1},w_{v2},w_{a1},w_{a2}\) from calculated Kalman gain, the recursive updates read

\begin{equation}
\label{1_6}
\begin{aligned}
{v_i} &= {w_{v1}}\frac{{{p_i} - {p_{i - 1}}}}{T} + {w_{v2}}{v_{i - 1}}\\
      &= {w_{v1}}\frac{{{p_i} - {p_{i - 1}}}}{T} \\
      &+ {w_{v1}}{w_{v2}}\frac{{{p_{i - 1}} - {p_{i - 2}}}}{T} + {w_{v1}}w_{v2}^{\rm{2}}\frac{{{p_{i - {\rm{2}}}} - {p_{i - {\rm{3}}}}}}{T}{\rm{ + }}...\\
      &= w_{v1}\sum_{k=0}^{i-1} w_{v2}^{k}\,
       \frac{p_{\,i-k}-p_{\,i-k-1}}{T},\\
{a_i} &= {w_{a1}}\frac{{{p_i} - 2{p_{i - 1}} + {p_{i - 2}}}}{{{T^2}}} + {w_{a2}}{a_{i - 1}}\\
      &= {w_{a1}}\frac{{{p_i} - 2{p_{i - 1}} + {p_{i - 2}}}}{{{T^2}}}\\
      &+ {w_{a1}}{w_{a2}}\frac{{{p_{i - 1}} - 2{p_{i - 2}} + {p_{i - 3}}}}{{{T^2}}} \\
      &+ {w_{a1}}w_{a2}^{\rm{2}}\frac{{{p_{i - 2}} - 2{p_{i - 3}} + {p_{i - 4}}}}{{{T^2}}}{\rm{ + }}...\\
      &= w_{a1}\sum_{k=0}^{i-1} w_{a2}^{k}\,
        \frac{p_{\,i-k}-2p_{\,i-k-1}+p_{\,i-k-2}}{T^{2}} .
\end{aligned}
\end{equation}

Given \(p_i,v_i,a_i\), an \(n\)-step-ahead position prediction follows from kinematics:

\begin{equation}
\label{1_7}
p_{i+n}=p_i+v_i\,nT+\tfrac12 a_i\,n^{2}T^{2}.
\end{equation}

Substituting~\eqref{1_6} into~\eqref{1_7} yields a weighted finite-difference expansion

\begin{equation}
\label{1_8}
\begin{aligned}
p_{i+n} &=\, p_i + nTw_{v1}\sum_{k=0}^{i-1} w_{v2}^{k} \frac{p_{\,i-k}-p_{\,i-k-1}}{T}\\
&+\tfrac12 n^{2}T^{2} w_{a1}\sum_{k=0}^{i-1} w_{a2}^{k}\frac{p_{\,i-k}-2p_{\,i-k-1}+p_{\,i-k-2}}{T^{2}}\\
&= \sum_{k=0}^{i} w_k p_k,
\end{aligned}
\end{equation}

Under model~\eqref{1_4}, long-term prediction ultimately reduces to a weighted sum of past positions, an intrinsic limitation when the target executes nonlinear manoeuvres.  To validate the potential of position-only estimated states, we construct the corresponding state-space models.

We begin with the \emph{two-position estimator} (E2P), based on  
\[
p_{i+1}=p_i+(p_i-p_{i-1}),
\]
which yields
\begin{equation}
\label{eq:E2P}
\begin{aligned}
\mathbf x_{i+1} &=
\begin{bmatrix}
2 & -1\\
1 &  0
\end{bmatrix}\mathbf x_i +
\begin{bmatrix}
1&0\\[2pt]
0&1
\end{bmatrix}\mathbf u_i, 
&
\\
\mathbf z_i &=
\begin{bmatrix}
1&0
\end{bmatrix}\mathbf x_i + \mathbf v_i,
\end{aligned}
\end{equation}
with \(\mathbf x_i=[p_i,p_{i-1}]^{\mathsf T}\).

\medskip
By extending the position history, we obtain higher-order variants.  Defining  
\(\mathbf x_i=[p_i,p_{i-1},p_{i-2}]^{\mathsf T}\) gives the \emph{three-position estimator} (E3P),  
and \(\mathbf x_i=[p_i,p_{i-1},p_{i-2},p_{i-3}]^{\mathsf T}\) yields the \emph{four-position estimator} (E4P):
\begin{equation}
\label{eq:E3P,E4P}
\begin{aligned}
p_{i+1}&=p_i+\tfrac12(p_i-p_{i-1})+\tfrac12(p_{i-1}-p_{i-2}),\\
p_{i+1}&=p_i+\tfrac13(p_i-p_{i-1})+\tfrac13(p_{i-1}-p_{i-2})\\
       &+\tfrac13(p_{i-2}-p_{i-3}).
\end{aligned}
\end{equation}

A \emph{variable-weight four-position estimator} (E4PVW) further refines the weights:
\begin{equation}
\label{eq:E4PVW}
\begin{aligned}
p_{i+1} &= p_i + \frac{3(p_i-p_{i-1})}{6}\\
&+ \frac{2(p_{i-1}-p_{i-2})}{6} + \frac{(p_{i-2}-p_{i-3})}{6}.
\end{aligned}
\end{equation}

\medskip
For even greater flexibility, the weights may be \emph{learned} from data.  The offline \emph{four-position estimator with regressed weights} (E4PRW) uses
\begin{equation}
\label{eq:E4PRW}
\begin{aligned}
p_{i+1}
=& 1.2668\,p_i - 0.0152\,p_{i-1}\\
&+ 0.0103\,p_{i-2} - 0.2618\,p_{i-3},
\end{aligned}
\end{equation}
while its online counterpart (E4PTRW) updates these coefficients at each step using the most recent 50 samples.

\begin{figure}[htbp]
  \centering
  \subfloat[Position prediction with different estimators.]{
    \includegraphics[width=.45\columnwidth]{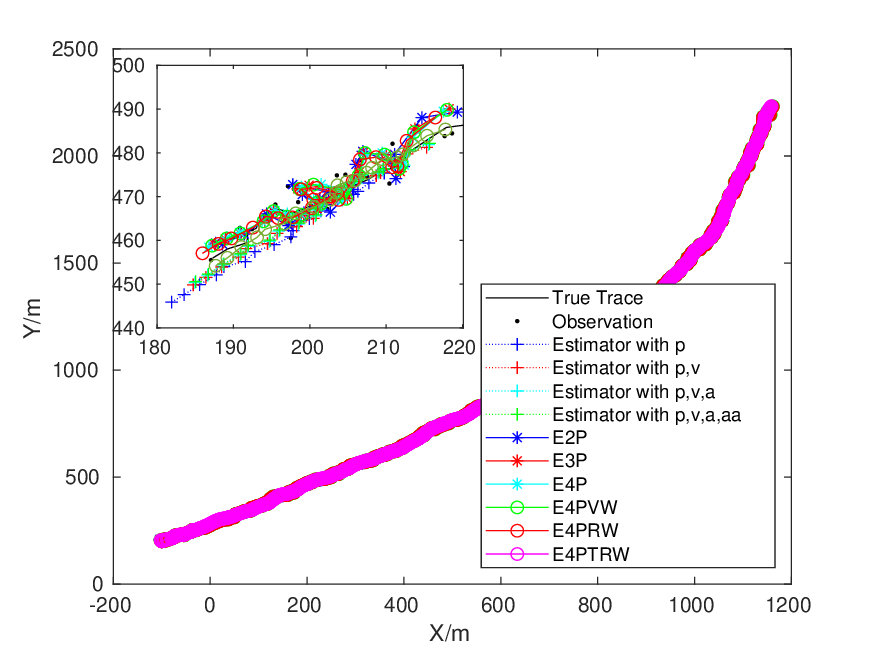}}
  \quad
  \subfloat[Accumulated prediction error.]{
    \includegraphics[width=.45\columnwidth]{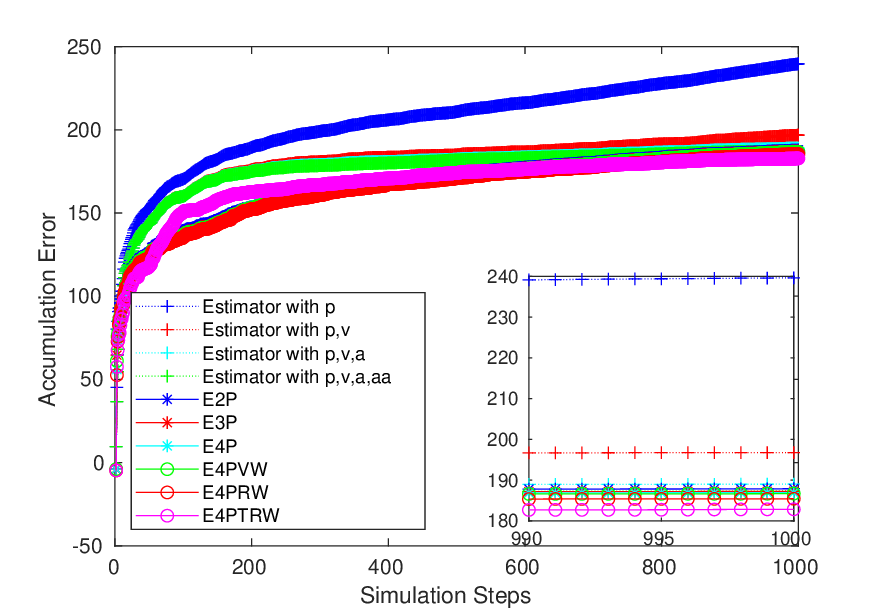}}
  \caption{Comparison of classical Kalman filters with position-stack estimators.}
  \label{fig:pos_pred}
\end{figure}

\medskip
Simulation results in Fig.~\ref{fig:pos_pred} show that all position-stack estimators outperform classical Kalman filters built on  
\([p_i]\), \([p_i,v_i]\), \([p_i,v_i,a_i]\), and \([p_i,v_i,a_i,\dot a_i]\) (the latter including jerk \(\dot a_i\)).  
Accuracy increases with longer stacks (E3P, E4P) and peaks for variants with adaptive or regressed weights (E4PVW, E4PRW, E4PTRW), with E4PRW notably narrowing the error band and E4PTRW further enhancing adaptability.

\medskip
Despite these gains, embedding a dedicated linear-regression step in each prediction is conceptually inelegant and risks overfitting when the target's motion pattern changes abruptly.

A more principled solution is to regard the regression coefficients themselves as \emph{latent states}, updating them in real time alongside the kinematic variables.  Accordingly, we extend the state vector to include not only these coefficients but, when appropriate, the full set of weights of a neural network approximation of the target dynamics.  These augmented states are then estimated online using a NNSSM.

\section{Neural Network State-Space Model}\label{sec3}

\begin{figure}[htbp]
  \centering
  \includegraphics[width=.9\columnwidth]{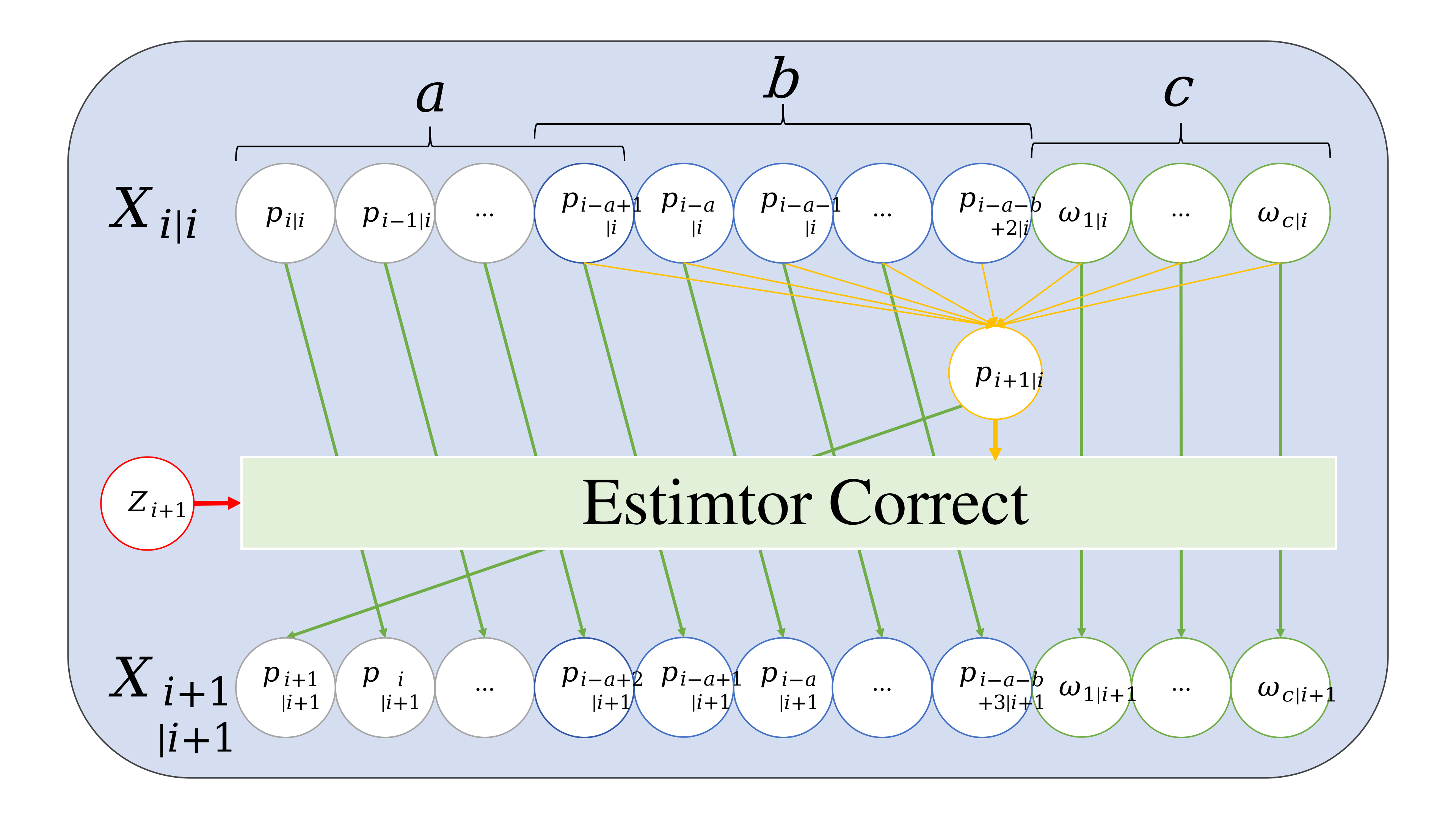}
  \caption{Neural network state-space model.}
  \label{fig:NNSSM}
\end{figure}

Based on the concept of treating neural network weights as estimated states, a new state-space model to achieve state estimation for maneuverable targets is introduced as \eqref{3_1} and Fig.~\ref{fig:NNSSM}.

\noindent

\begin{equation}
\begin{aligned}
\mathbf x_{i+1}
  &= f_i(\mathbf x_i,\mathbf u_i) = f_i\bigl(\bigl[\,p_i,\dots,p_{i-a+1},\\
  &\quad\;\;p_{i-a-b+2},w_{1},\dots,w_{c}\bigr]^{\!\mathsf T},\,\mathbf u_i\bigr)\\
  &= \bigl[\,p_{i+1},p_i,\dots,p_{i-a-b+3},w_{1},\dots,w_{c}\bigr]^{\!\mathsf T}
     + \mathbf u_i\\
\mathbf z_i
  &= h_i(\mathbf x_i,\mathbf v_i)
   = p_i + \mathbf v_i
\end{aligned}
\label{3_1}
\end{equation}

In the proposed model, the state vector comprises \((a-1)+b+c\) elements in total, namely \(a+b-1\) historical position states and \(c\) weight parameters. Here:
\begin{itemize}
  \item \(a\) is the prediction horizon in discrete steps.  For example, if the sampling interval is \(T=0.01\)s and a 0.1s ahead prediction is required, then \(a = 10\);
  \item \(b\) is the number of input-layer nodes of the neural network being simulated;
  \item \(c\) is the total number of weight parameters in that network.
\end{itemize}
These \((a-1)+b+c\) states are jointly estimated online via the NNSSM.

Let

\noindent
\begin{equation}
\begin{aligned}
\mathbf P_{i,[n:m]}\;\triangleq & \;\bigl[p_{i-n},\,p_{i-n-1},\,\dots,\,p_{i-m}\bigr]\\
\mathbf W\;\triangleq & \;\bigl[w_{1},\,w_{2},\,\dots,\,w_{c}\bigr],
\label{3_2_1}
\end{aligned}
\end{equation}

and denote by \(f_{\mathrm{NN}}(\cdot)\) the neural network mapping from the previous state to the new predicted position \(p_{i+1}\).  A more detailed state-space model can then be written as
\noindent
\begin{equation}
\begin{aligned}
\mathbf x_{i+1} =& f_i(\mathbf x_i,\mathbf u_i) \\ 
        =& \bigl[f_{\mathrm{NN}}\bigl(\mathbf P_{i,\,[(a-1):(a+b-2)]},\,\mathbf W\bigr),\\
         & \mathbf P_{i,\,[0:(a+b-3)]},\;\mathbf W\bigr]^{\!\mathsf T} + \mathbf u_i \\ 
        =& [\,p_{i+1},\;\mathbf P_{i,\,[0:(a+b-3)]},\;\mathbf W\,]^{\!\mathsf T},\\
\mathbf z_i     =& p_i + \mathbf v_i.
\end{aligned}
\label{eq:NNSSM_detailed}
\end{equation}

The a-frame-ahead prediction function is therefore given by
\begin{equation}
\label{eq:NNSSM_pre}
p_{i+a} \;=\; f_{\mathrm{NN}}\bigl(\mathbf P_{i,\,[0:(b-1)]},\,\mathbf W\bigr).
\end{equation}

\begin{figure}[htbp]
  \centering
  \includegraphics[width=.9\columnwidth]{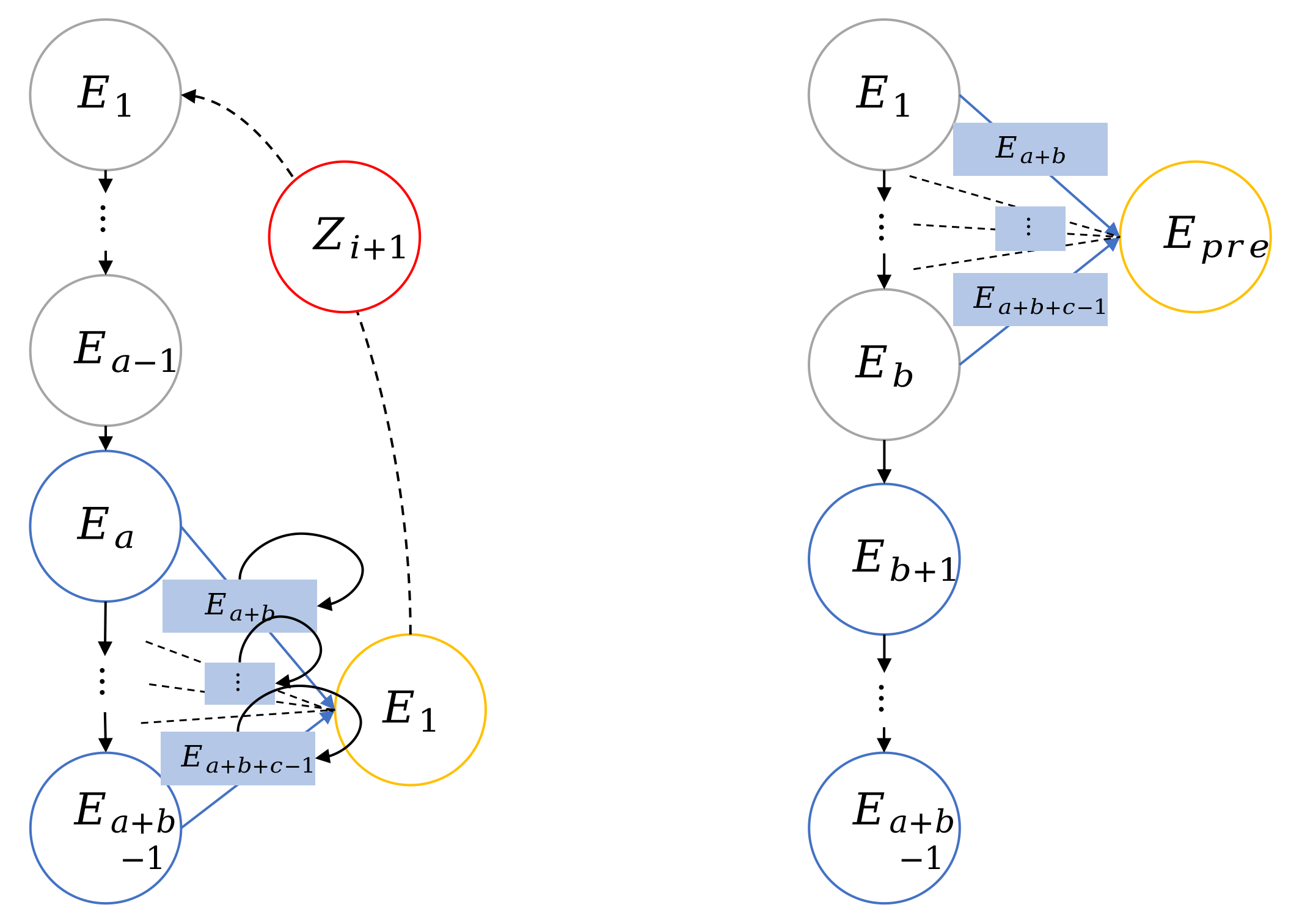}
  \caption{Simplest neural network state-space model: weighted-sum iteration.}
  \label{fig:SNNSSM}
\end{figure}

Fig.~\ref{fig:SNNSSM} illustrates the iterative structure of the simplest neural network—a single weighted-sum node—whose architecture defines the specific form of \(f_{\mathrm{NN}}(\cdot)\).  Let \(E_j\) denote the \(j\)-th component of the estimated state vector \(\mathbf x_i\).  The corresponding state-space representation is  
\begin{equation}
\label{eq:NNSSM_simple}
\begin{aligned}
&\mathbf x_{i+1} = f_i\bigl(\mathbf x_i,\mathbf u_i\bigr) \\[2pt]
        &= \Bigl[\mathbf P_{i,[(a-1):(a+b-2)]}\,\mathbf W^{\!\mathsf T},\;
                 \mathbf P_{i,[0:(a+b-3)]},\;
                 \mathbf W\Bigr]^{\!\mathsf T} + \mathbf u_i \\[2pt]
        &= \bigl[p_{i+1},\;\mathbf P_{i,[0:a+b-3]},\;\mathbf W\bigr]^{\!\mathsf T},\\[4pt]
&\mathbf z_i     = p_i + \mathbf v_i.
\end{aligned}
\end{equation}

The needed position prediction formula is  
\begin{equation}
\label{eq:NNSSM_simple_pre}
\begin{aligned}
E_{\mathrm{pre}} &= p_{i+a} = f_{\mathrm{NN}}\bigl(\mathbf P_{i,\,[0:(b-1)]},\,\mathbf W\bigr)\\
   &= \mathbf P_{i,[0:(b-1)]}\,\mathbf W^{\!\mathsf T}.
\end{aligned}
\end{equation}

The weighted sum in~\eqref{eq:NNSSM_simple} furnishes the \emph{prior} one-step prediction \(E_{1}\) in
\(\mathbf x_{i+1|i}\). After the new observation \(\mathbf z_{i+1}\) is received, the estimator updates every state component \(E_j\) to obtain the \emph{posterior} estimate \(\mathbf x_{i+1|i+1}\).

For an \(a\)-frame look-ahead, the algorithm~\eqref{eq:NNSSM_simple_pre} uses the most recent \(b\) position estimates  
\(E_{1},\dots,E_{b}\) and the current weight vector \(E_{a+b},\dots,E_{a+b+c-1}\) into the same weighted-sum relation, producing the required multi-frame forecast \(E_{pre}\).

\begin{figure}[htbp]
  \centering
  \includegraphics[width=1\columnwidth]{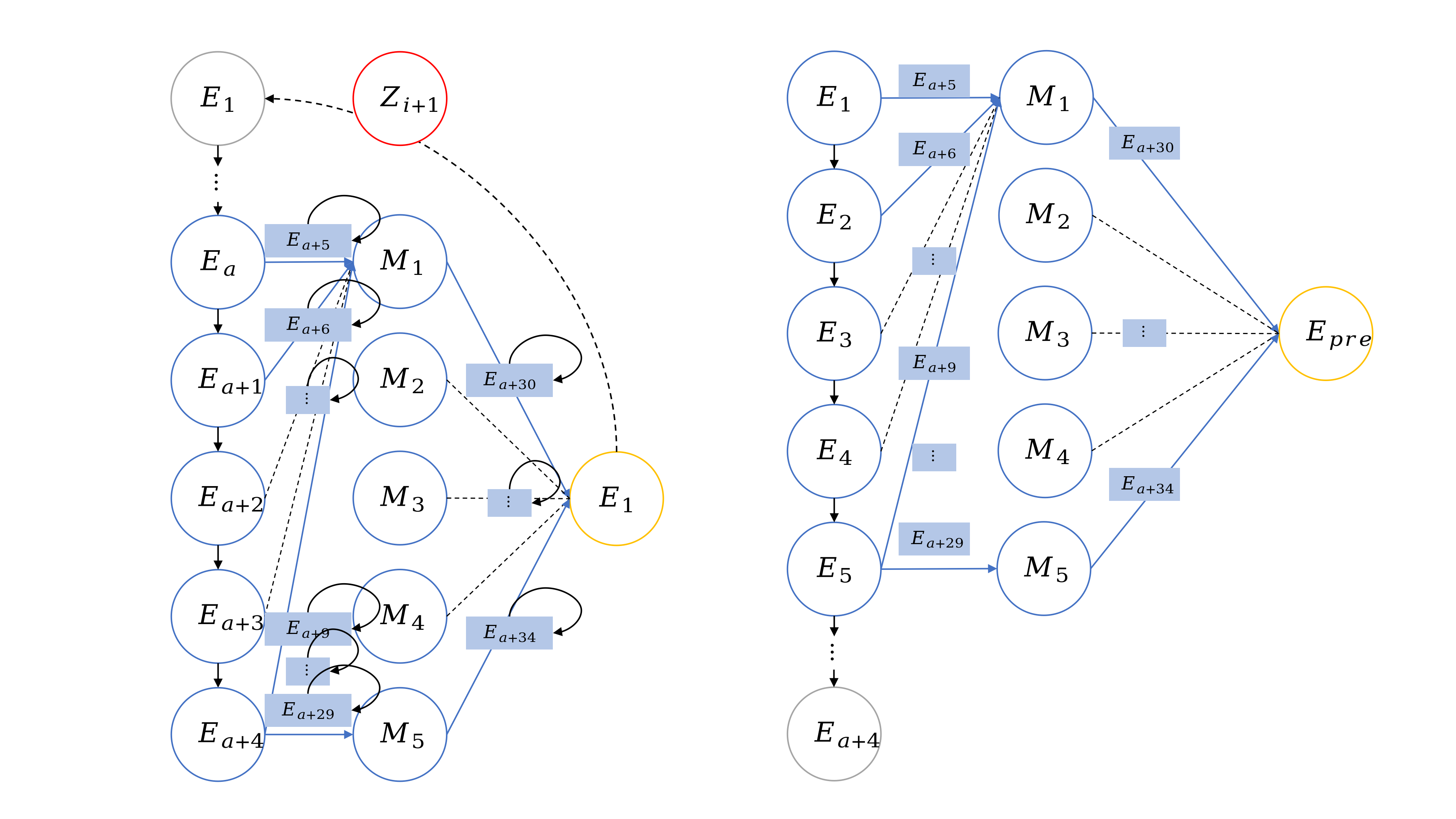}
  \caption{Simple neural network state-space model: 5-5-1 architecture.}
  \label{fig:NNSSM2}
\end{figure}

Although the weighted-sum model provides a conceptual bridge to neural networks, it lacks the representational richness of even a modest multi-layer perceptron such as the 5-5-1 architecture shown in Fig.~\ref{fig:NNSSM2}, which depicts a simple fully connected feedforward network with five input nodes, one hidden layer of five neurons, and a single output node (a “5-5-1” architecture).  

To perform an \(a\)-frame-ahead prediction, the augmented state vector must include \(a+4\) past positions plus the network's weight parameters, for a total of \(a+34\) estimated states.  Here, \(M_j\) denotes the activation of the \(j\)th hidden neuron.  The mapping \(f_{\mathrm{NN}}(\cdot)\) in~\eqref{eq:NNSSM_detailed} is a nonlinear function implemented by a multi-layer recurrent architecture.

The precise layer sizes, connectivity pattern, and choice of activation functions should be determined based on the target application and available computational resources.

\section{Implementation of the Neural Network State-Space Model}\label{sec4}

Having formulated the NNSSM, we now describe its implementation with a classical Unscented Kalman Estimator, hereafter abbreviated \emph{NNSSE-UKE}.

\subsubsection*{1.\;Parameter Initialization}

For an $n$-dimensional state $\hat{\mathbf{x}}_{i|i}$ choose the spread $\alpha$, the secondary scaling $\beta$, and the scaling factor $\kappa$.  
Define
\begin{equation}
\lambda = \alpha^{2}\,(n+\kappa)-n. \label{NNSSE-UKE_1}
\end{equation}
Initialise the state $\hat{\mathbf{x}}_{0}$, the a-posteriori covariance $\mathbf \Pi_{0\mid0}$, and the noise covariances $\mathbf Q$ (process) and $\mathbf R$ (measurement).

\subsubsection*{2.\;Sigma-Point Generation (Posterior State)}

With $\hat{\mathbf{x}}_{i|i}$ and $\mathbf \Pi_{i\mid i}$ form $2n+1$ sigma points:
\begin{equation}
\begin{aligned}
&\mathbf X_{0}=\hat{\mathbf{x}}_{i\mid i},\\
&\mathbf X_{l}=\hat{\mathbf{x}}_{i\mid i}\pm\bigl(\sqrt{(n+\lambda)\,\mathbf \Pi_{i\mid i}}\bigr)_{l},\quad
l=1,\dots,n.
\end{aligned}\label{NNSSE-UKE_2}
\end{equation}
Weights for the state mean and covariance are
\begin{equation}
\begin{aligned}
w^{(0)}_{S}&=\tfrac{\lambda}{n+\lambda}, &
w^{(0)}_{\Pi}&=w^{(0)}_{S}+1-\alpha^{2}-\beta,\\
w^{(l)}_{S}&=w^{(l)}_{\Pi}=\tfrac{1}{2(n+\lambda)}, & l&=1,\dots,2n. 
\end{aligned}\label{NNSSE-UKE_3}
\end{equation}

\subsubsection*{3.\;Propagation through the State Equation}

Propagate every sigma point through $f_i(\cdot)$ (see Eq.~\eqref{eq:NNSSM_detailed}):
\begin{equation}
\hat{\mathbf X}_{l}=f_i\bigl(\mathbf X_{l}\bigr),\qquad l=0,\dots,2n. \label{NNSSE-UKE_5}
\end{equation}

\subsubsection*{4.\;Time Update}

\begin{equation}
\begin{aligned}
\hat{\mathbf x}_{i+1\mid i} &= \sum_{l=0}^{2n} w_{S}^{(l)}\,\hat{\mathbf X}_{l},\\
\Pi_{i+1\mid i}    &= \sum_{l=0}^{2n} w_{\Pi}^{(l)}\bigl(\hat{\mathbf X}_{l}-\hat{\mathbf x}_{i+1\mid i}\bigr)
                     \bigl(\hat{\mathbf X}_{l}-\hat{\mathbf x}_{i+1\mid i}\bigr)^{\!\mathsf T}+\mathbf Q.
\end{aligned}\label{NNSSE-UKE_6}
\end{equation}

\subsubsection*{5.\;Sigma-Point Generation (Prior State) and Observation}

Generate $\tilde{\mathbf X}_{l}$ from $\hat{\mathbf x}_{i+1\mid i},\mathbf \Pi_{i+1\mid i}$ using ~ \eqref{NNSSE-UKE_2};
pass them through the measurement model $h(\cdot)$:
\begin{equation}
\mathbf Z_{l}=h\bigl(\tilde{\mathbf X}_{l}\bigr). \label{NNSSE-UKE_8}
\end{equation}
Compute
\begin{equation}
\begin{aligned}
\hat{\mathbf z}_{i+1} &= \sum_{l=0}^{2n} w_{S}^{(l)} {\mathbf Z}_{l},\\
\mathbf \Pi_{zz} &= \sum_{l=0}^{2n} w_{\mathbf \Pi}^{(l)}\bigl({\mathbf Z}_{l}-\hat{\mathbf z}_{i+1}\bigr)\bigl({\mathbf Z}_{l}-\hat{\mathbf z}_{i+1}\bigr)^{\!\mathsf T}+{\mathbf R},\\
\mathbf \Pi_{xz} &= \sum_{l=0}^{2n} w_{\mathbf \Pi}^{(l)}\bigl(\tilde {\mathbf X}_{l}-\hat{\mathbf x}_{i+1\mid i}\bigr)\bigl({\mathbf Z}_{l}-\hat{\mathbf Z}_{i+1}\bigr)^{\!\mathsf T}.
\end{aligned}\label{NNSSE-UKE_9}
\end{equation}

\subsubsection*{6.\;Measurement Update}

\begin{equation}
\begin{aligned}
&\mathbf K = \mathbf \Pi_{xz}\,\mathbf \Pi_{zz}^{-1},\\
&\hat{\mathbf x}_{i+1\mid i+1}= \hat{\mathbf x}_{i+1\mid i}+\mathbf K\bigl(\mathbf z_{i+1}-\hat{\mathbf z}_{i+1}\bigr),\\
&\mathbf \Pi_{i+1\mid i+1}= \mathbf \Pi_{i+1\mid i}-\mathbf K\mathbf \Pi_{zz}\mathbf K^{\mathsf T}. 
\end{aligned}\label{NNSSE-UKE_10}
\end{equation}

\subsubsection*{7.\;Position Prediction}

The updated state \(\hat{\mathbf x}_{i+1\mid i+1}\) yields the position forecast \(\hat p_{i+a}\) ~ \eqref{eq:NNSSM_pre} and the algorithm returns to Step~2 for the next iteration.  

\section{Simulation and Experiment}\label{SecSimAndExp}

The proposed NNSSM can be instantiated with diverse nonlinear estimators.  In this section, we demonstrate its generality by constructing three NNSSEs based on representative EKE, UKE, and PE. By choosing prediction horizon \(a=3\), input dimension \(b=25\), and weight count \(c=25\), we obtain the variants \emph{NNSSE-EKE}, \emph{NNSSE-UKE}, and \emph{NNSSE-PE}, all derived from \eqref{eq:NNSSM_simple}.

To illustrate NNSSM's architectural flexibility, we fix \(a=3\) and employ UKE to realise estimators for different network topologies:
\begin{itemize}
  \item \emph{NNSSE-5-5-1}: 5 input nodes, one hidden layer of 5 neurons, and one output node (see Fig.~\ref{fig:NNSSM2});
  \item \emph{NNSSE-10-10-1}: 10 input nodes, one hidden layer of 10 neurons, and one output node;
  \item \emph{NNSSE-Tanh}: the 5-5-1 architecture with Tanh activations in the hidden layer;
  \item \emph{NNSSE-5-5-5-1}: 5 input nodes, two hidden layers of 5 neurons each, and one output node.
\end{itemize}
These configurations demonstrate the algorithm's ability to simulate and estimate a wide range of neural network structures.

In our experimental study, we also train established neural models—including \emph{RNN}, \emph{LSTM}, \emph{GRU}, \emph{TCN}, \emph{NeuralODE}, \emph{Transformer}, and an \emph{Online Transformer} (trained adaptively at runtime)—to benchmark the adaptability and rapid learning capability of NNSSE against these baselines. All of these networks employ an input layer of 25 historical position values and between two and four hidden layers; their precise architectures are detailed in the code repository accompanying this paper.

\subsection{Simulation}\label{SecSim}

\begin{figure}[htbp]
  \centering
  \subfloat[True vs.\ estimated trajectories over the full simulation horizon.]{
  \includegraphics[width=0.44\columnwidth]{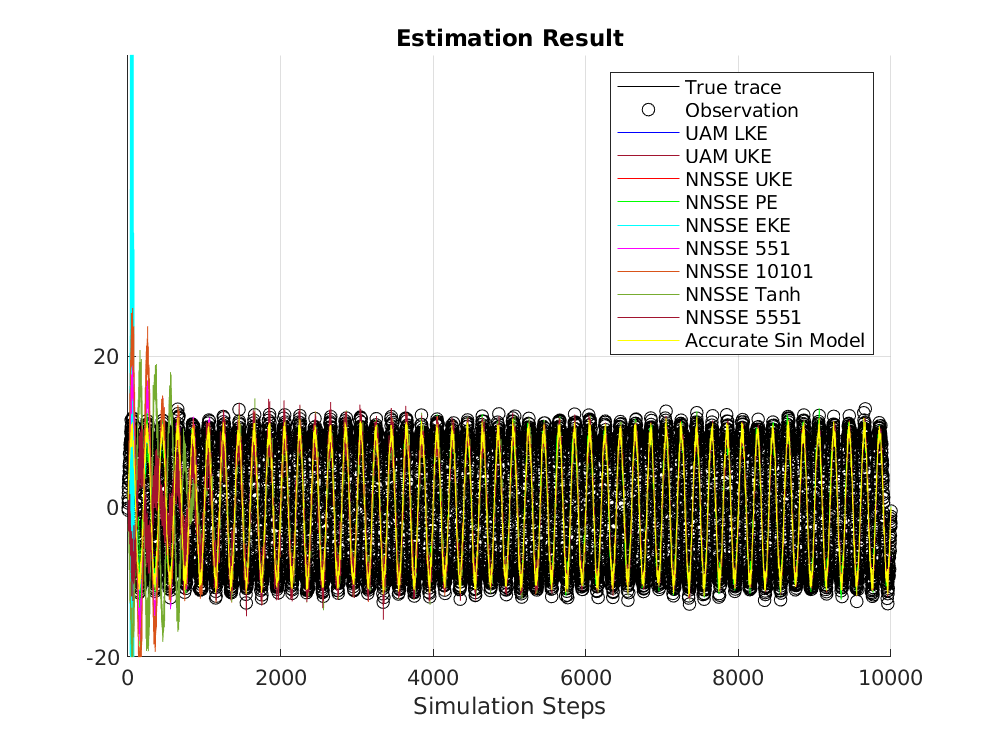}
  \label{fig:sim_traj1}
  }
  \quad
  \subfloat[Estimator predictions during the midle phase.]{
  \includegraphics[width=0.44\columnwidth]{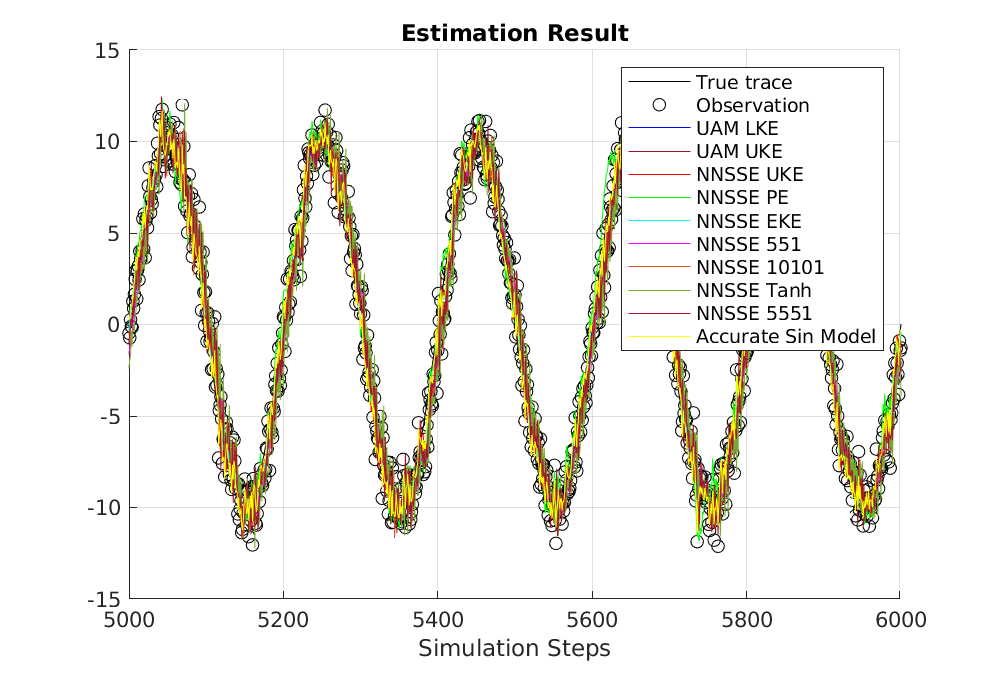}
  \label{fig:sim_traj2}
  }
  \caption{Comparison of classical Kalman filters with position-stack estimators.}
  \label{fig:sim_trajs}
\end{figure}

A noisy sine wave is used to emulate a manoeuvring target.  
The sensor sampling rate is \(200~\mathrm{Hz}\) with a latency of \(0.15~\mathrm{s}\) (three-frame prediction);  
the ground-truth motion is a \(1~\mathrm{s}\) sine with amplitude \(10\), and the observation noise is zero-mean Gaussian with unit covariance. 

For comparison, a uniformly accelerated model (UAM) is estimated by both a linear Kalman estimator (\emph{UAM-LKE}) and UKE (\emph{UAM-UKE}). An \emph{Accurate-Sin-Model} reference uses the exact sine model and LKE.

Fig.~\ref{fig:sim_traj1}-\ref{fig:sim_traj2} show the trajectories and estimation errors.  
\emph{NNSSE-EKE} exhibits large transients in the first few dozen steps, whereas \emph{NNSSE-10-10-1} requires more data to converge owing to its larger parameter set.

\begin{figure}[htbp]
  \centering
  \includegraphics[width=.9\columnwidth]{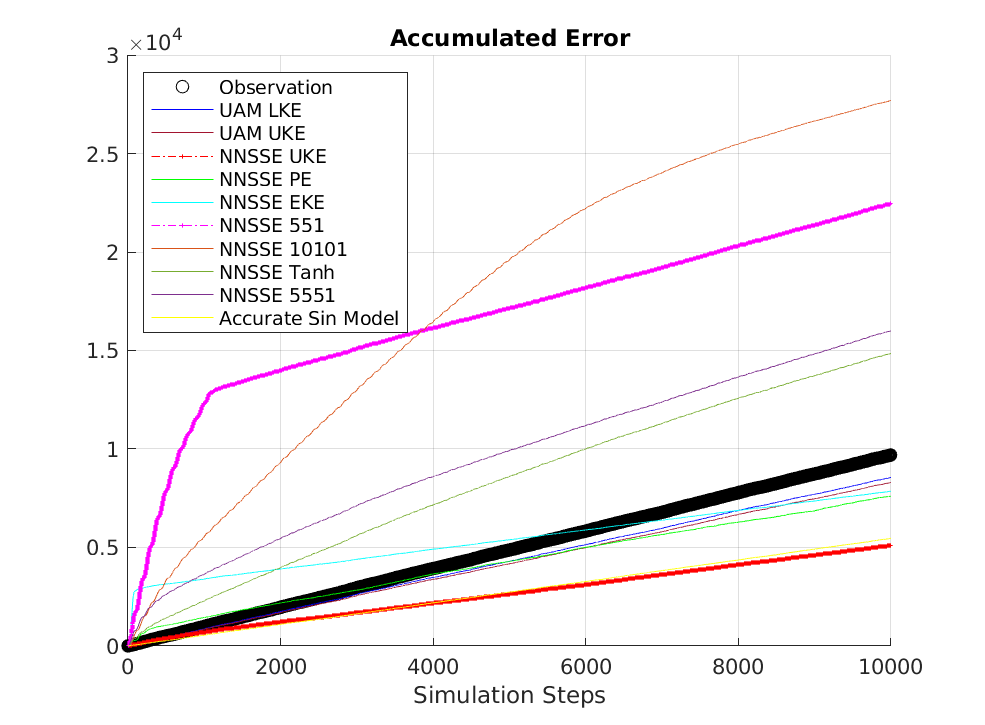}
  \caption{Accumulated prediction error for all estimators over 10000 steps.}
  \label{fig:sim_err}
\end{figure}

\begin{table}[h]
\caption{Accumulated error ($\times10^{4}$) and run time for 10000 steps.  
Maximum values are \textcolor{red}{\textbf{red}}, minima \textcolor{green}{\textbf{green}}.}
\label{tab:error_time}
\centering
\begin{tabular}{@{}lccc@{}}
\toprule
Estimator & Steps 1-10000 & Steps 8000-10000 & Time (s) \\ 
\midrule
UAM-LKE               & 0.8554 & 0.1681 & 0.0718 \\ 
UAM-UKE               & 0.8303 & 0.1633 & 0.2908 \\ 
Accurate-Sin (ideal)  & 0.5465 & 0.1105 & 0.0687 \\ 
\cmidrule(lr){2-4}
NNSSE-UKE             & \textcolor{green}{\textbf{0.5104}} & 0.0985 & 7.0647 \\ 
NNSSE-PE              & 0.7577 & 0.1327 & 22.8345 \\ 
NNSSE-EKE             & 0.7852 & \textcolor{green}{\textbf{0.0975}} & \textcolor{green}{\textbf{0.5203}} \\ 
NNSSE-5-5-1           & 2.2451 & 0.2137 & 6.7968 \\ 
NNSSE-10-10-1         & \textcolor{red}{\textbf{2.7693}} & 0.2209 & \textcolor{red}{\textbf{35.0905}} \\ 
NNSSE-Tanh            & 1.4850 & 0.2280 & 7.2935 \\ 
NNSSE-5-5-5-1         & 1.6003 & \textcolor{red}{\textbf{0.2353}} & 16.6581 \\ 
\bottomrule 
\end{tabular}
\end{table}

Fig.~\ref{fig:sim_err} plots the error after step~200, when all NNSSEs reach steady performance. \emph{NNSSE-UKE}, \emph{NNSSE-PE}, and \emph{NNSSE-EKE} all surpass their UAM-based counterparts and even outperform the \emph{Accurate-Sin-Model} reference despite having no model priors.

Table~\ref{tab:error_time} summarises accumulated error and computation time.  
NNSSEs incur higher cost because of the enlarged state, yet the longest run (35.09 s for 10000 steps) remains practical.  
Among them, \emph{NNSSE-EKE} is the most economical but suffers from early instability and Jacobian requirements, making UKE the preferred implementation.
  
\subsection{Dual-Mirror Tracking Experiment}\label{SecMirrorExp}

\begin{figure}[htbp]
  \centering
  \subfloat[Optical layout]{\includegraphics[width=.45\columnwidth]{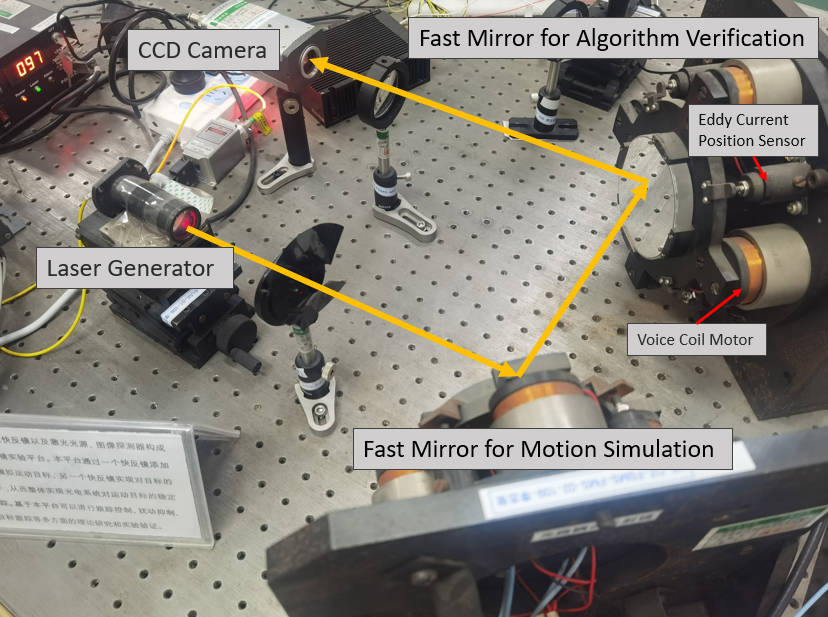}}\hfill
  \subfloat[Mechanical assembly]{\includegraphics[width=.45\columnwidth]{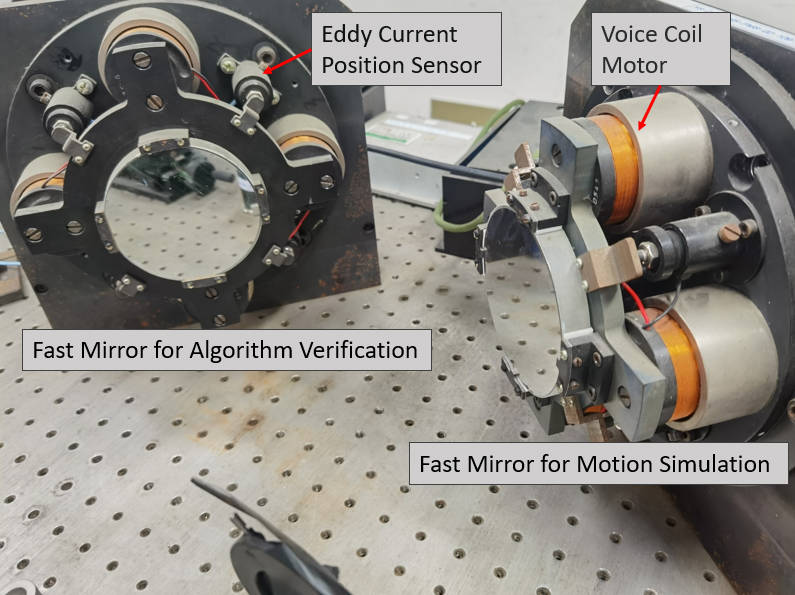}}
  \caption{Dual fast mirror experimental platform.}
  \label{fig:mirror_platform}
\end{figure}

\begin{figure}[htbp]
  \centering
  \includegraphics[width=.9\columnwidth]{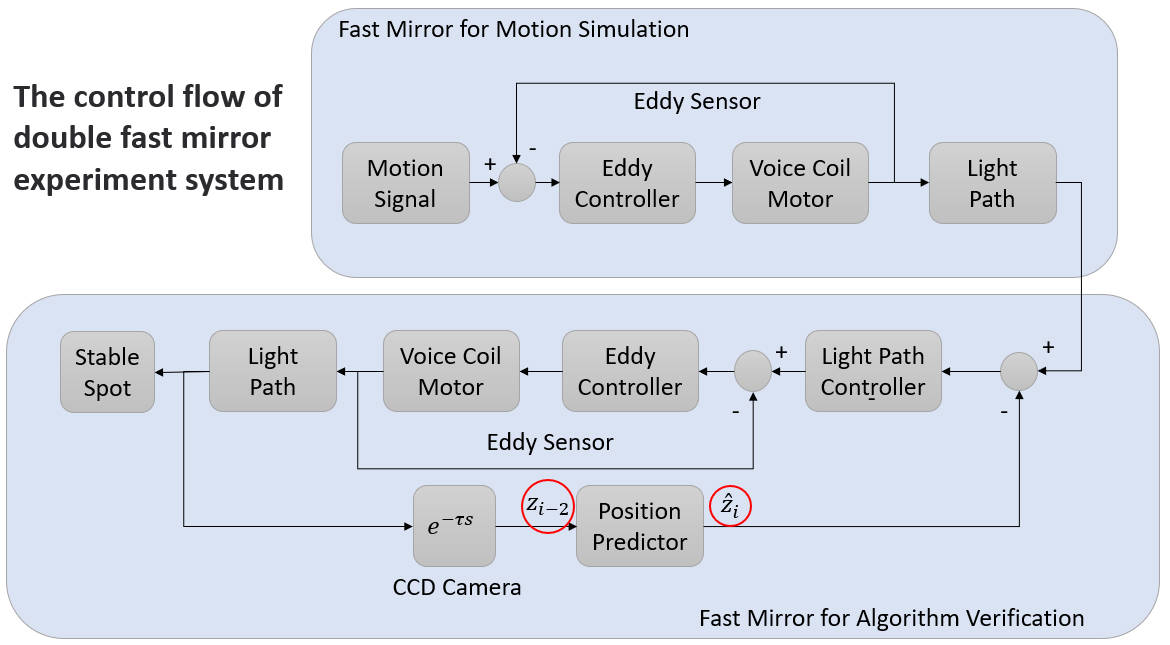}
  \caption{Control flow of the dual-reflection mirror platform. A three-frame prediction compensates for the CCD latency.}
  \label{fig:mirror_control}
\end{figure}

The proposed NNSSEs were further validated on a dual-reflection mirror platform.  
As depicted in Fig.~\ref{fig:mirror_platform}, the bench comprises a laser source, a CCD camera, a fast steering mirror that emulates target motion, and a validation mirror for closed-loop testing.  
Each steering mirror is driven by four voice-coil actuators and monitored by four eddy-current displacement sensors.

The control objective is to maintain the laser spot at the centre of the CCD.  
Although the camera samples at \(200\;\mathrm{Hz}\), a \(0.015~\mathrm{s}\) latency exists, so a three-frame-ahead prediction is inserted in the feedback loop (Fig.~\ref{fig:mirror_control}).  
Extensive position sequences were first collected to train all baseline neural networks. A new sequences from the same source were then fed to both the trained networks and the NNSSEs for a fair comparison.

\begin{figure}[h]
  \centering
  \subfloat[Full record.]{\includegraphics[width=.45\columnwidth]{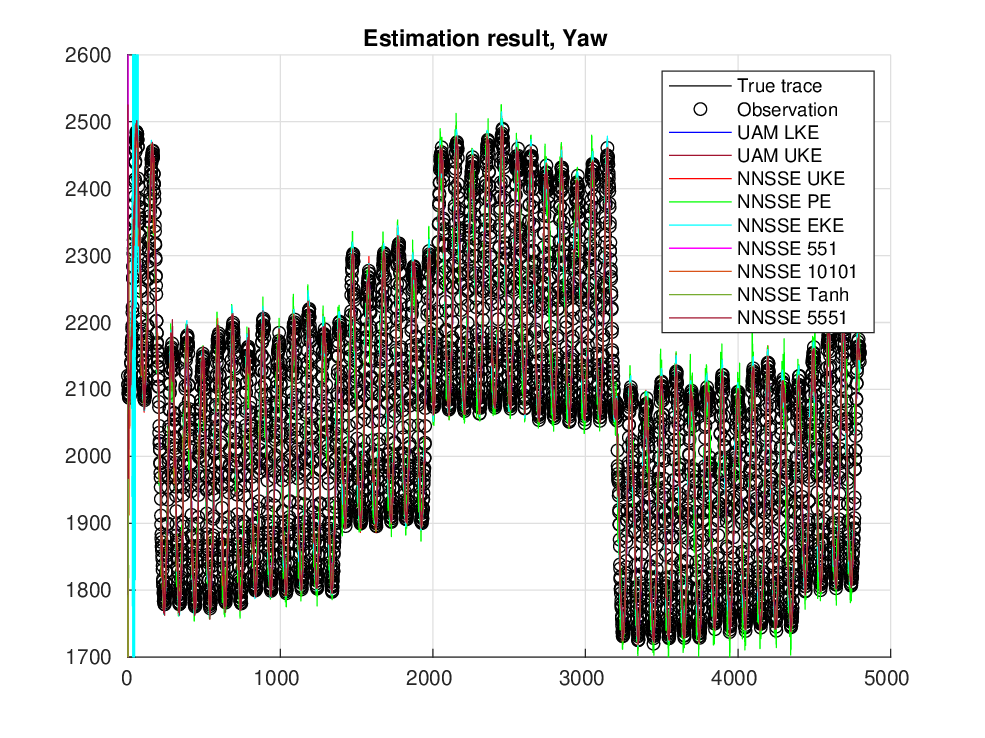}}\hfill
  \subfloat[Enlarged central segment.]{\includegraphics[width=.45\columnwidth]{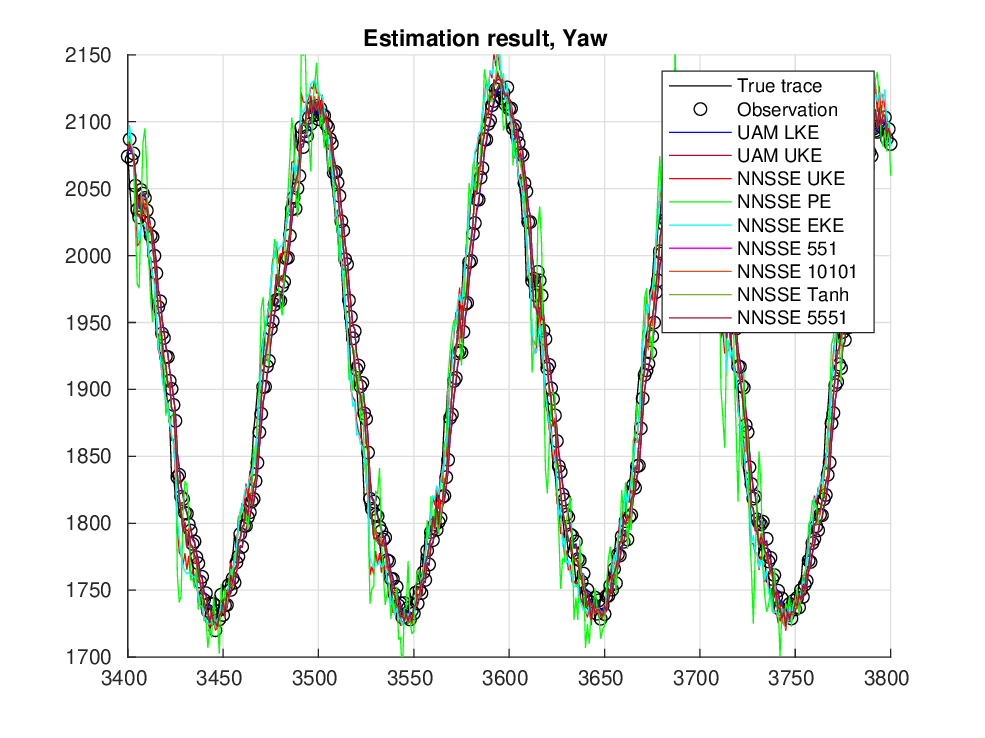}}
  \caption{Spot trajectory and NNSSE estimates.}
  \label{fig:Y0}
\end{figure}

\paragraph{Overall tracking performance.}
Fig.~\ref{fig:Y0} shows the measured spot trajectory together with the estimates produced by the different NNSSEs; the right panel magnifies the central segment.

\begin{figure}[h]
  \centering
  \includegraphics[width=.9\columnwidth]{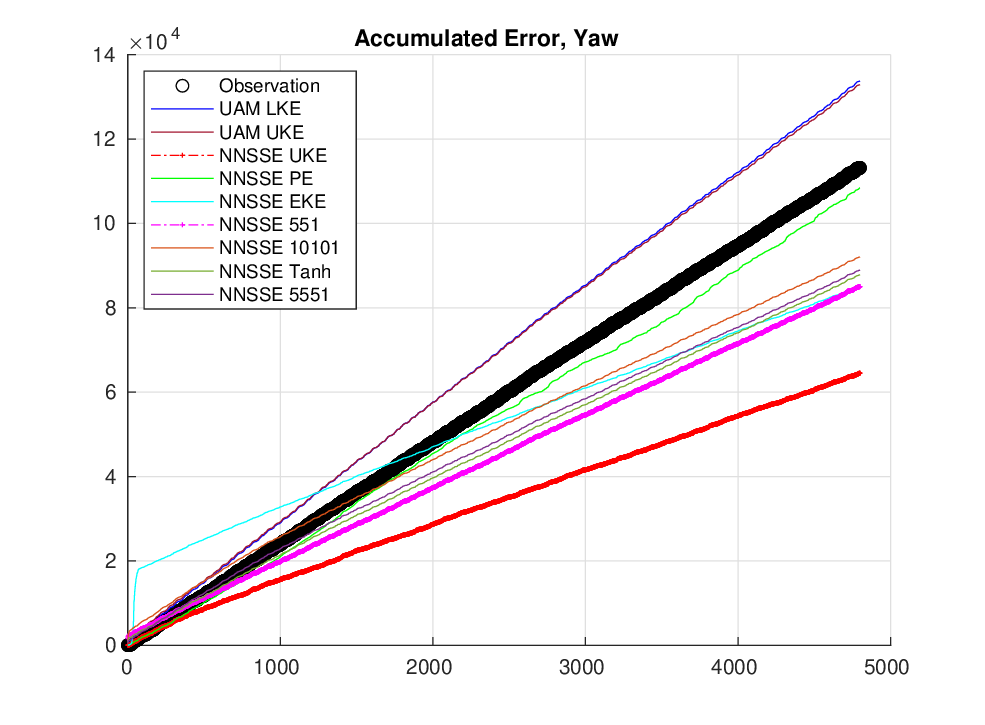}
  \caption{Cumulative tracking error of competing estimators.}
  \label{fig:Y1}
\end{figure}

Cumulative-error curves are plotted in Fig.~\ref{fig:Y1}.  
Without artificially injected Gaussian noise in simulation, every NNSSE surpasses its UAM-based counterpart; quantitative results are summarised in Table~\ref{tab:yaw_acc}.  

\begin{table}[h]
\caption{Dual-mirror experiment: cumulative error (\(\times10^{5}\)).  
Maximum values are \textcolor{red}{\textbf{red}}, minima \textcolor{green}{\textbf{green}}.}
\label{tab:yaw_acc}
\centering
\begin{tabular}{@{}lccc@{}}
\toprule
Estimator & Steps 1-4799 & Steps 2000-4799 & Time (s) \\
\midrule
UAM-LKE              & 1.3374 & 0.7614 & 0.0357 \\
UAM-UKE              & 1.3289 & 0.7533 & 0.1375 \\
\cmidrule(lr){2-4}
NNSSE-UKE            & \textcolor{green}{\textbf{0.6450}} & \textcolor{green}{\textbf{0.3585}} & 3.5825 \\
NNSSE-PE             & \textcolor{red}{\textbf{1.0842}} & \textcolor{red}{\textbf{0.6297}} & 10.3143 \\
NNSSE-EKE            & 0.8509 & 0.3802 & \textcolor{green}{\textbf{0.2624}} \\
NNSSE-5-5-1          & 0.8504 & 0.4768 & 3.3630 \\
NNSSE-10-10-1        & 0.9206 & 0.4805 & \textcolor{red}{\textbf{16.3013}} \\
NNSSE-Tanh           & 0.8781 & 0.4812 & 3.4833 \\
NNSSE-5-5-5-1        & 0.8892 & 0.4783 & 7.7643 \\
\bottomrule 
\end{tabular}
\end{table}

\begin{figure}[htbp]
  \centering
  \subfloat[Limited training]{\includegraphics[width=0.45\columnwidth]{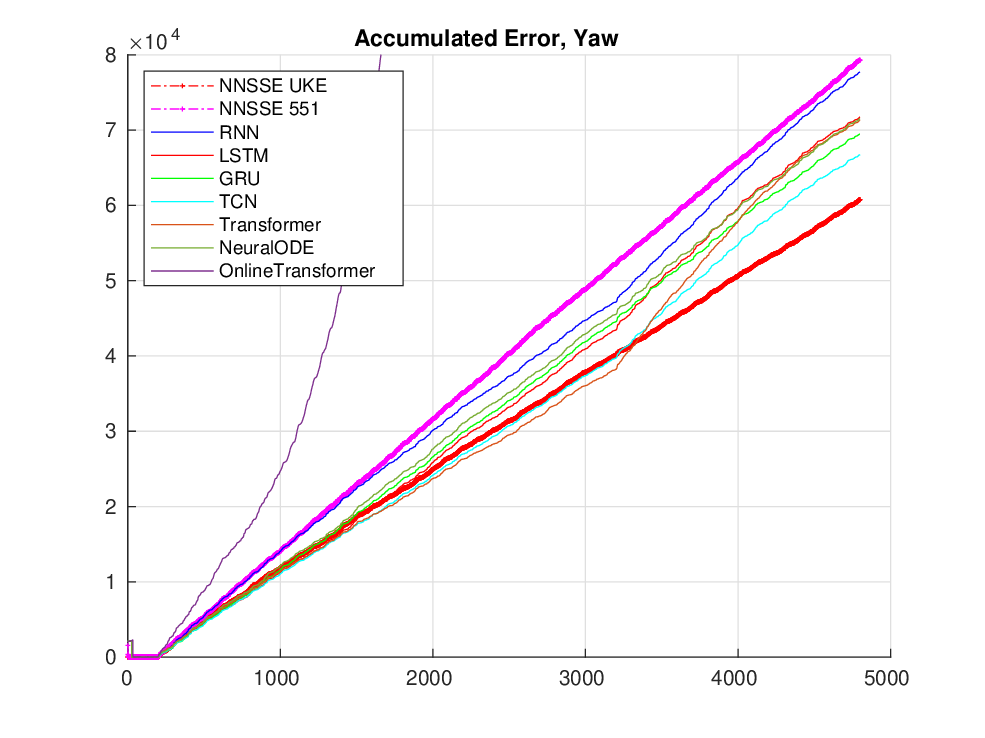}}\hfill
  \subfloat[Full training]{\includegraphics[width=0.45\columnwidth]{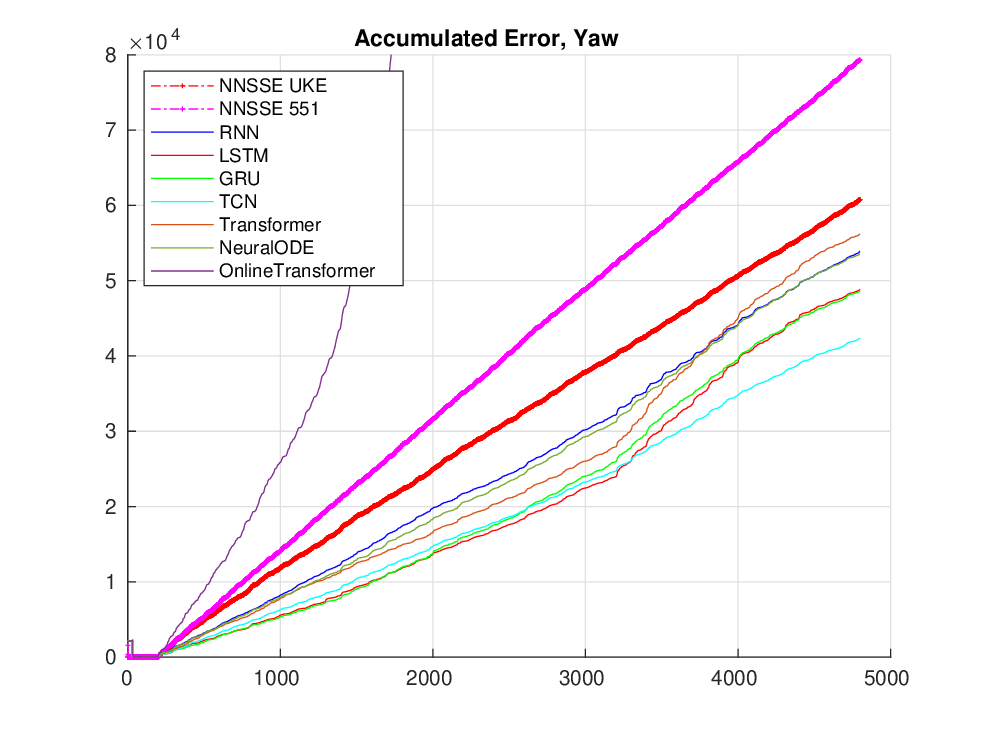}}
  \caption{Cumulative prediction error of neural baselines under (a) limited and (b) full training regimes.}
  \label{fig:Y2}
\end{figure}

\paragraph{Comparison with neural baselines.}
We evaluate two training regimes: (i) limited training (10 epochs for most networks; 50 epochs for the Transformer) and (ii) full training (200+ epochs on 10000 samples). Fig.~\ref{fig:Y2}(a) and Fig.~\ref{fig:Y2}(b) show the cumulative prediction error under each regime, and Table ~\ref{tab:nn_acc} summarizes the error measured from step 200 onward.

\begin{table*}[h]
\caption{Cumulative error (\(\times10^{5}\)) for limited and full training.
Maximum values are \textcolor{red}{\textbf{red}}, minima \textcolor{green}{\textbf{green}}.}
\label{tab:nn_acc}
\centering
\begin{tabular*}{\textwidth}{@{\extracolsep\fill}lcccc}
\toprule
& \multicolumn{2}{c}{Limited Training} & \multicolumn{2}{c}{Full Training} \\
\cmidrule(lr){2-3}\cmidrule(lr){4-5}
Method                & \shortstack{Steps 200--4799} & \shortstack{Steps 2000--4799} 
                      & \shortstack{Steps 200--4799} & \shortstack{Steps 2000--4799} \\
\midrule
NNSSE-UKE             & 0.6076 & 0.3585 & 0.6076 & 0.3585 \\
NNSSE-5-5-1           & 0.7929 & 0.4768 & 0.7929 & 0.4768 \\
\cmidrule(lr){2-5}
RNN                   & 0.7770 & 0.4759 & 0.5392 & 0.3416 \\
LSTM                  & 0.7172 & 0.4588 & 0.4882 & 0.3521 \\
GRU                   & 0.6947 & 0.4289 & 0.4858 & 0.3487 \\
TCN                   & \textcolor{green}{\textbf{0.6673}} & \textcolor{green}{\textbf{0.4259}} & \textcolor{green}{\textbf{0.4237}} & \textcolor{green}{\textbf{0.2771}} \\
Transformer           & 0.7137 & 0.4768 & 0.5617 & 0.3968 \\
NeuralODE             & 0.7152 & 0.4390 & 0.5356 & 0.3527 \\
Online Transformer    & \textcolor{red}{\textbf{4.6542}} & \textcolor{red}{\textbf{3.3732}} & \textcolor{red}{\textbf{4.6246}} & \textcolor{red}{\textbf{3.4425}} \\
\bottomrule 
\end{tabular*}
\end{table*}

The NNSSEs attain prediction accuracy comparable to that of the neural baselines trained for only a few epochs.  
Although they lag behind the fully-trained networks in the interval from step 200 to step 4799,  
all NNSSE variants converge rapidly: after roughly 2000 learning steps their performance rivals that of the fully-trained models, falling significantly short only of the TCN.

\begin{figure}[htbp]
  \centering
  \subfloat[Spot trajectory and estimation]{\includegraphics[width=0.45\columnwidth]{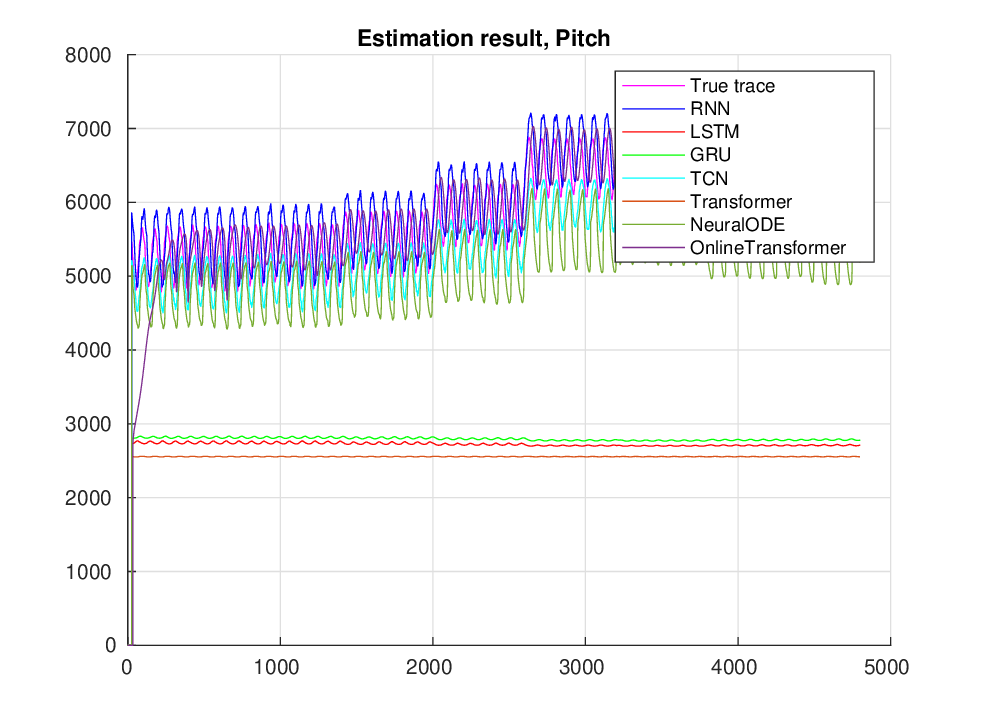}}\hfill
  \subfloat[Cumulative prediction error of neural baseline]{\includegraphics[width=0.45\columnwidth]{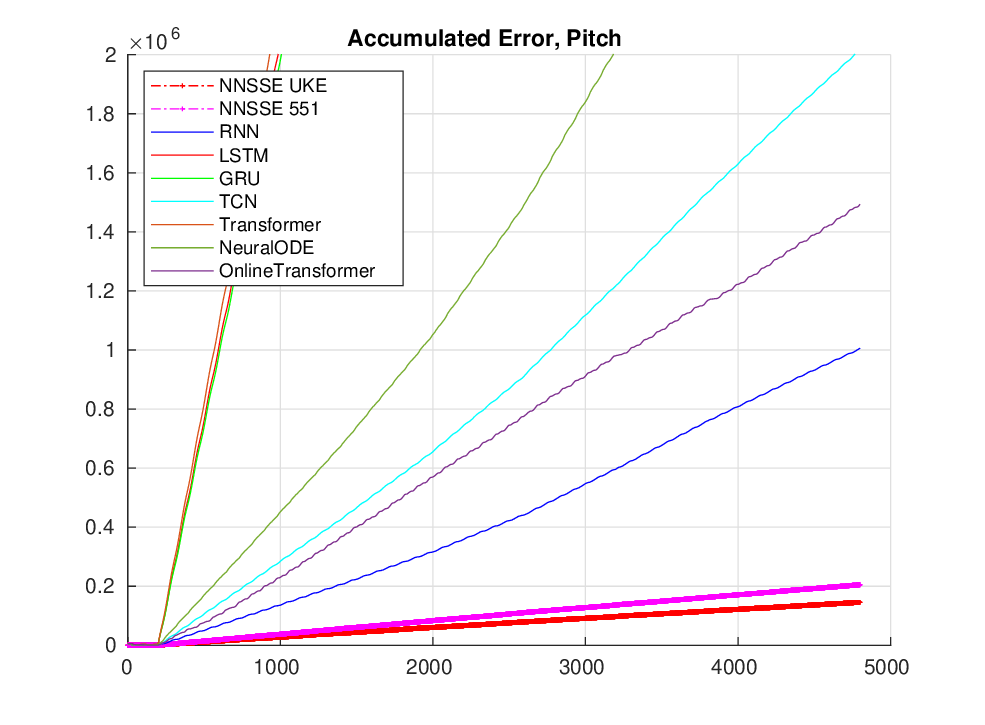}}
  \caption{Estiamte pitch trajectory using yaw-trained networks.}
  \label{fig:DR_P}
\end{figure}

To further test generalisation, we applied the yaw-trained networks to a new data set in the pitch axis.  
The pretrained RNN, LSTM, GRU, TCN, NeuralODE, Transformer, OnlineTransformer were unable to deliver useful forecasts.
By contrast, all NNSSEs quickly adapted to the new motion pattern and maintained high accuracy.  
Fig.~\ref{fig:DR_P}(a) shows the pitch trajectory and estimation result, and Fig.~\ref{fig:DR_P}(b) plots the corresponding cumulative errors.

\subsection{Drone-Tracking Experiment}\label{SecDroneExp}

The efficacy of the proposed approach was further evaluated on a real unmanned aerial vehicle trajectory.  

\begin{figure}[htbp]
  \centering
    \includegraphics[width=.45\columnwidth]{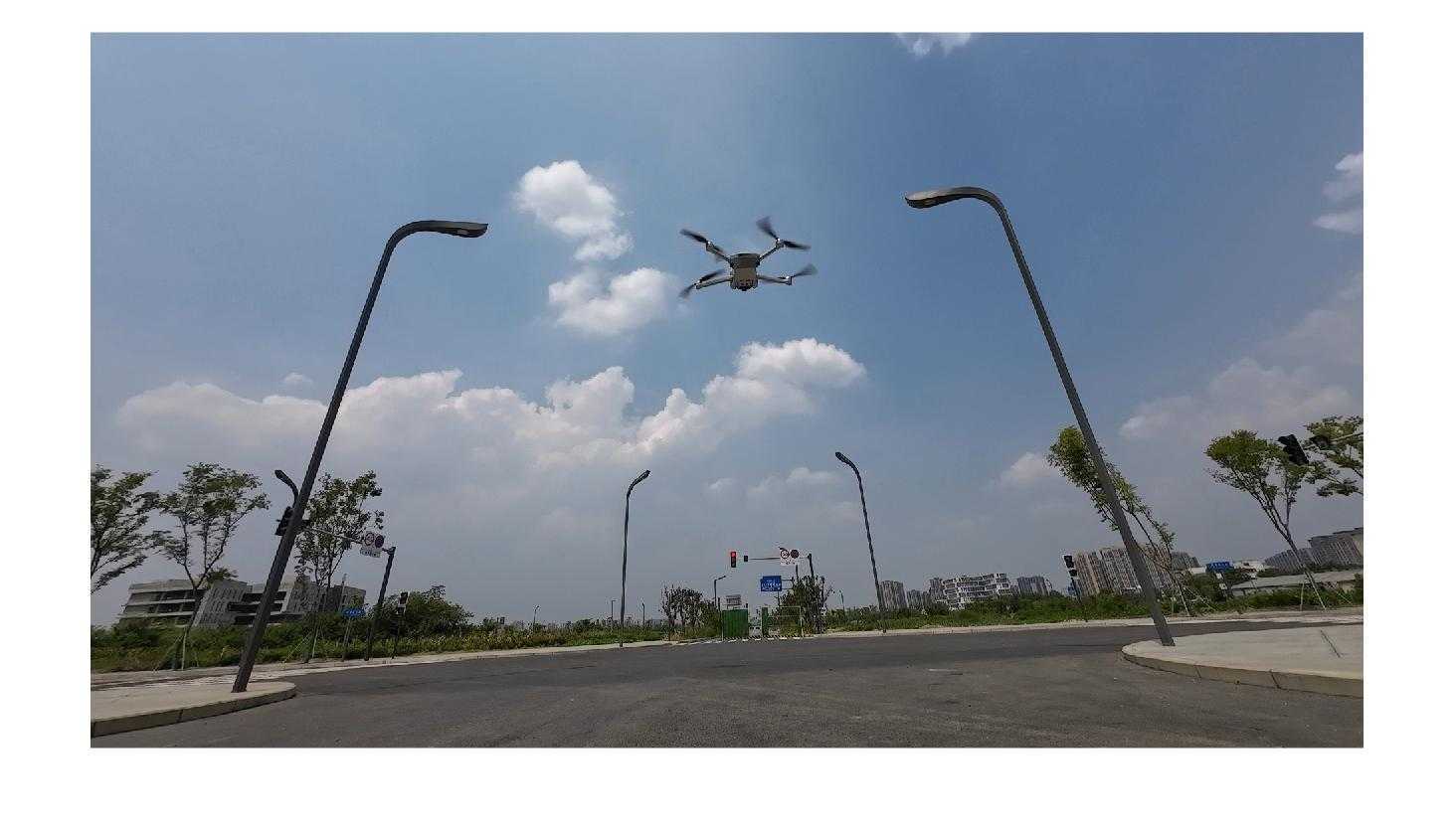}\hfill
    \includegraphics[width=.45\columnwidth]{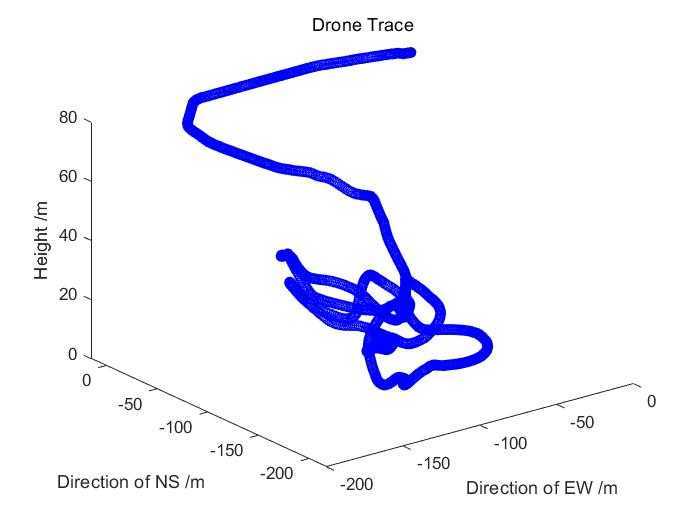}
  \caption{Drone and part of its trajectory.}
  \label{fig:Drone_0}
\end{figure}

\begin{figure}[htbp]
  \centering
  \subfloat[Measured and estimated yaw trajectories]{
    \includegraphics[width=.45\columnwidth]{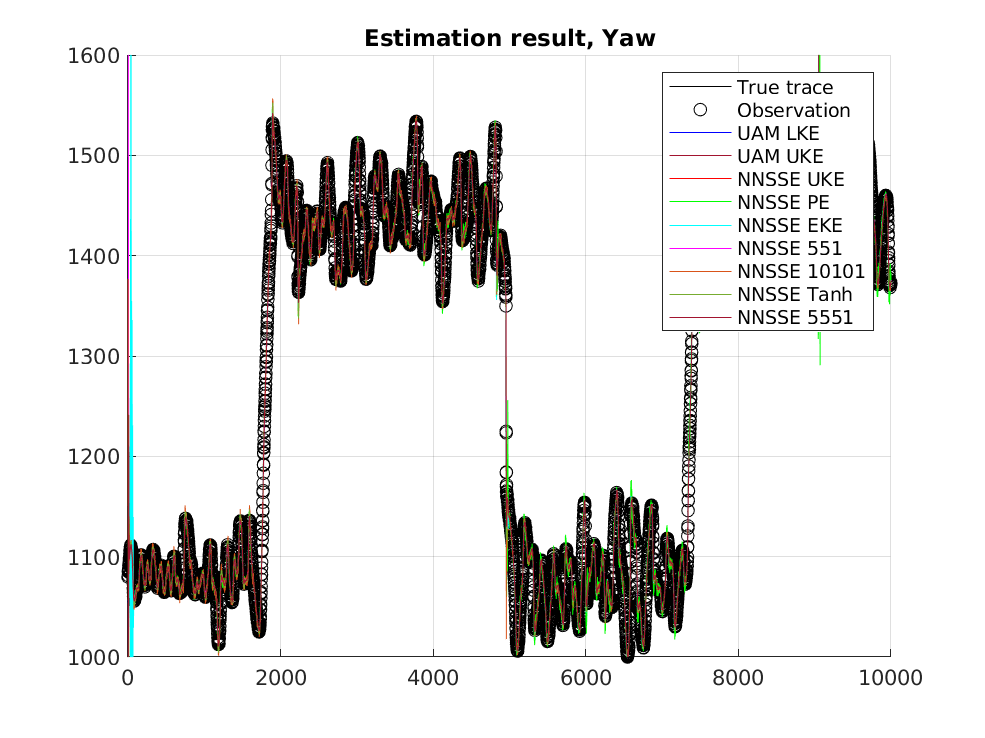}}\hfill
  \subfloat[Cumulative error of NNSSE variants]{
    \includegraphics[width=.45\columnwidth]{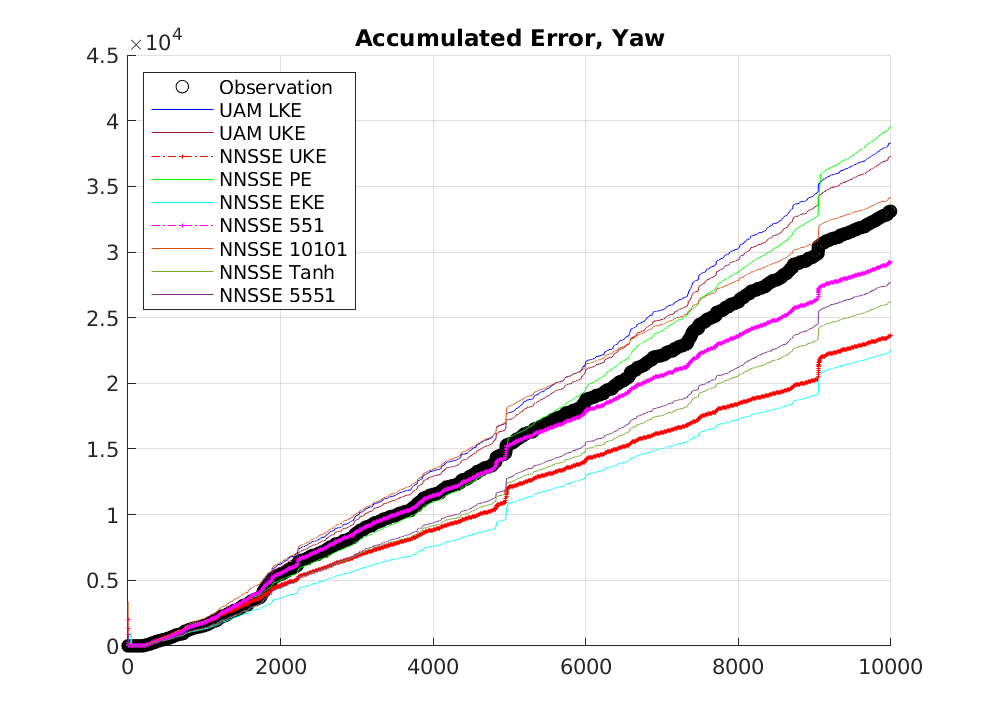}}
  \caption{Performance of NNSSEs on the UAV yaw data.}
  \label{fig:Drone_1}
\end{figure}

Fig.~\ref{fig:Drone_1}(a) plots the yaw trajectory together with the estimates delivered by the various NNSSEs.  
Unlike the quasi-sinusoidal laboratory path, the UAV executes complex, non-cooperative manoeuvres;  
the sharp 180$^{\circ}$ reversals arise from the \emph{zenith-crossing} phenomenon (i.e., the target passes directly overhead and the yaw changes sign).  
Fig.~\ref{fig:Drone_1}(b) shows the corresponding cumulative errors: after a short learning phase, most NNSSEs attains a markedly lower error than the UAM methods baseline.

\begin{figure}[htbp]
  \centering
  \subfloat[Fully-trained neural predictions]{%
    \includegraphics[width=.45\columnwidth]{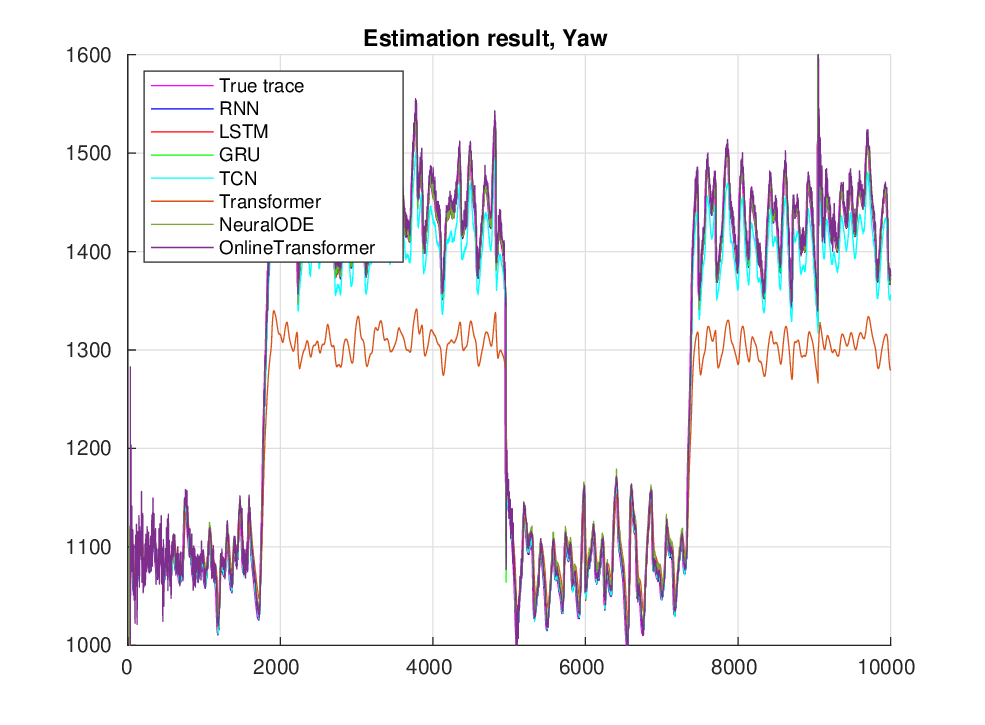}}\hfill
  \subfloat[Cumulative error of neural baselines]{%
    \includegraphics[width=.45\columnwidth]{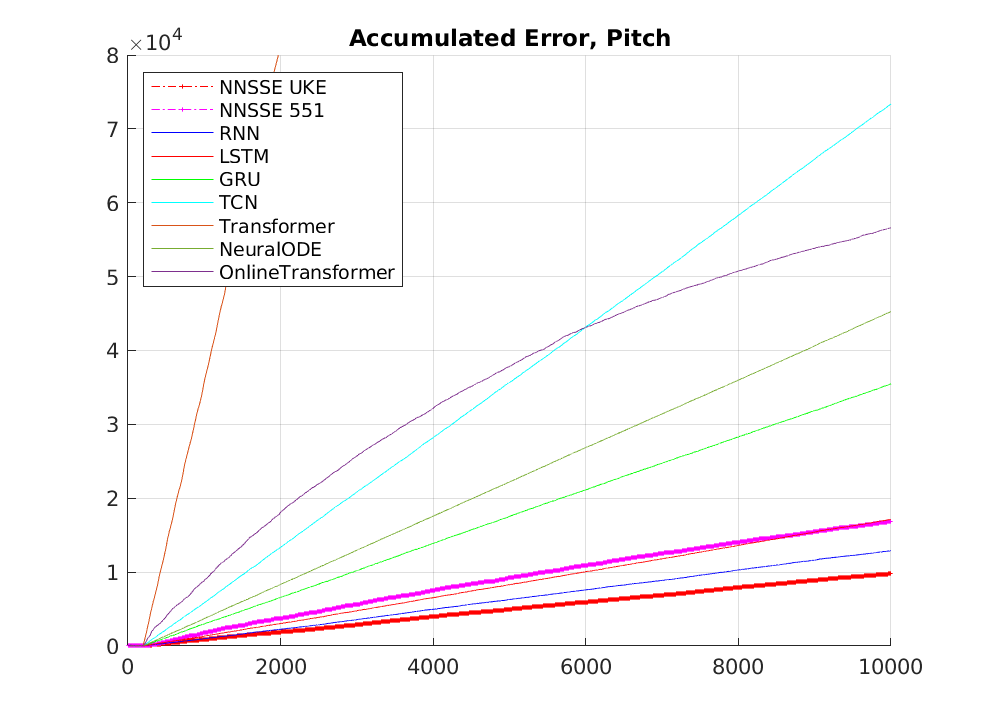}}
  \caption{Neural baselines on the UAV yaw data.}
  \label{fig:Drone_1}
\end{figure}

For a fair comparison with data-driven baselines, we trained every neural model on 20000 historical samples.  
Fig.~\ref{fig:Drone_1}(a) presents the predictions obtained after full training, and Fig.~\ref{fig:Drone_1}(b) reports the resulting cumulative errors.

Benefiting from the rapid-learning capability of the NNSSM, \emph{NNSSE-UKE} surpasses every pre-trained neural network in prediction accuracy.  
We anticipate that further optimisation of the network architecture and hyper-parameters will yield even higher precision.

\section{Conclusion}\label{Conclusion}

This paper has addressed the problem of tracking non-cooperative targets by first validating the feasibility of using a pure position sequence as the state vector in a classical state-space formulation.  
Building on this result, we introduced the \emph{Neural Network State-Space Model}, which needn't target's state-space model and treats both the target's recent positions and every weight of a surrogate neural network as latent states.  
The resulting nonlinear model can be realised with different nonlinear estimator; in extensive trials, UKE provided the best overall performance, and its full algorithmic structure has been presented in detail.

To demonstrate generality, NNSSM was further instantiated with EKE, UKE, and PE, producing three baseline NNSSEs that outperform constant-acceleration Kalman filters on noisy sine trajectories, a dual-reflection mirror platform, and real UAV flights.  
Additional experiments showed that NNSSM can emulate a variety of network topologies (5-5-1, 10-10-1, 5-5-5-1, Tanh activations, etc.) simply by adjusting the augmented state.  
When compared with fully pre-trained RNN, LSTM, GRU, TCN, NeuralODE, Transformer, and Online-Transformer baselines, the proposed NNSSEs achieved comparable—or, after a short on-line learning phase, superior—accuracy while retaining the rapid adaptation capability of recursive estimators.

All source code, data and more result are openly available at  
\url{https://github.com/ShineMinxing/PaperNNSSE.git}, facilitating reproduction and further development.

\textbf{Limitations and outlook:}
\begin{enumerate}
  \item \textbf{Computational complexity.}  
        The Kalman update scales as \(O(n^{3})\) with the number of augmented states; simulating very deep or wide networks therefore becomes costly.  Efficient sparsification or reduced-rank techniques are worth investigating.
  \item \textbf{Model design.}  
        The current NNSSM uses only fully connected layers and simple activations.  Borrowing advanced architectural motifs (e.g.\ residual connections or attention) from modern deep learning may improve accuracy without enlarging the state excessively.
  \item \textbf{Estimator diversity.}  
        We evaluated the three foundational filters (EKE, UKE, PE).  Incorporating more recent Bayesian filters or adaptive variants may yield further gains, particularly for high-dimensional weight vectors.
\end{enumerate}

We believe that NNSSM enlarges the design space of state estimation by seamlessly blending data-driven neural representations with principled Bayesian filtering, thereby offering a practical alternative when explicit nonlinear modelling is intractable.

\bibliographystyle{IEEEtran}
\bibliography{NNSSE}

\end{document}